**Different dynamics of information transfer in depression revealed: an EEG study**

Milena Čukić[1]*, Slavoljub Radenković[2], Miodrag Stokić[3,4], and Danka Savić[5]

[1]Instituto de Tecnología del Conocimiento, Universidad Complutense de Madrid, Spain.

[2]TomTom, Amsterdam, the Netherlands;

[3]Institute for Experimental Phonetics and Speech Pathology, Belgrade, Serbia

[4]Research and Development Institute – Life Activities Advancement Center, Belgrade, Serbia

[5]Vinča Institute for Nuclear Physics, Laboratory of Theoretical and Condensed Matter Physics 020/2, University of Belgrade, Belgrade, Serbia.

* Correspondence concerning this article should be addressed to Milena Čukić, Ph.D., Koningin Wilhelminaplein 644, 1062KS Amsterdam, the Netherlands, tel:+31615178926 email: micukic@ucm.es or micu@3ega.nl

**Abstract**

Depression is a serious world health issue and many avenues of research are aiming at elucidating the mechanisms behind it. Recent findings confirm the importance of a disrupted functional connectivity within the fronto-limbic system and other candidate areas important for depression. The question behind our work is whether areas with confirmed aberrated functioning in Major Depressive Disorder (MDD) are actually involved in the network which has different dynamics from a healthy one.

On a sample of 21 depressed patients (11 women and 9 men) and 20 age-matched healthy controls (10 women and 10 men), we applied Transfer Entropy (TE) to quantify the directed dynamical interactions in the resting-state electroencephalographic (EEG) data recorded in our previous research in which we compared physiological complexity features of recurrently depressed patients and healthy controls. The dynamics of healthy resting-state EEG is substantially different from the dynamics of MDD brain: the interactions (information transfers) in healthy controls are numerous during resting state, contrary to MDD brains which are repeatedly showing the "isolated" activity in frontal, parietal and temporal areas. To the best of our knowledge, this is the first time that a graphical representation of information transfer and its directions is presented showing the differences between MDD and healthy controls. The BINNUE approach provided us with both influence and directions of influence between compared time series (epoch extracted from recorded EEG)

**Introduction**

In his 1948 'Cybernetics and Psychopathology', Norbert Wiener stated that … '(there is) …*nothing surprising in considering the functional mental disorders as fundamentally diseases of the memory, of the circulating information kept by the brain in the active state, and of the long-time permeability of synapses*' which is in line with recent research results in recurrent dynamics and functional integration of the brain. Wiener (1948) called depression, paranoia, and schizophrenia a "functional mental disorders". The recent findings of the physiological, structural and functional mechanisms underlying depression confirmed the importance of disrupted functional connectivity within fronto-limbic system in depression (Bluhm et al., 2009; Berman et al., 2011; Vederine et al., 2010; Zhang et al., 2011; de Kwaasteniet et al., 2013; Kim et al., 2013; Chen et al., 2017).

It is well known how serious the problem of depression is for the healthcare system and society as a whole (Mathers & Loncar, 2006; Gillan & Daw, 2017; WHO 2017; World Economic Forum 2019).

A number of studies reported aberrant connectivity in depression. In their fMRI study of medication-free patients with major depressive disorder (MDD), Grimm et al. (2007) showed the existence of hypoactivity in the left dorsolateral prefrontal cortex (DLPFC) and hyperactivity in the right DLPFC. In a more recent fMRI study, Ge et al. (2019) confirmed that the decreased connectivity of the right intermediate hippocampus (RIH) with the limbic regions was a distinguishing feature for treatment-resistant depression (Ge et al., 2019). On the other hand, several connectivity studies reported that functional connectivity exists between subgenual anterior cingulate cortex (ACC) and medial temporal lobe (MTL) in depression, as well as in hippocampus and amygdala (Mathews et al., 2008; Pezawas et al., 2005); Furman et al (2011) reported the frontostriatal functional connectivity in major depressive disorder (MDD), and Horn (2010) reported the correlation between functional connectivity of pregenual anterior cingulate cortex (pgACC) and severity of anhedonia in MDD. Bluhm examined the resting state default-mode network connectivity in early depression using a seed region of interest analysis (Bluhm et al., 2009) and confirmed decreased connectivity within the caudate nucleus. Their study showed significantly reduced correlation between precuneus/posterior cingulate cortex and the bilateral caudate in depression compared with controls. Berman examined connectivity of the default network specifically in the subgenual cingulate both on- and off-task, and also the relationship between connectivity and rumination in MDD (Berman et al., 2011). Their results showed characteristic higher functional neural connectivity between posterior cingulate cortex and subgenual cingulate cortex, but *during rest periods only*. Vederine et al. (2011) and Kwaasteniet et al. (2013) elaborated on abnormal functional connectivity in the fronto-limbic system. Using the combination of fMRI and functional anisotropy (FA), de Kwaasteniet confirmed that white matter integrity of the uncinate fasciculus was reduced, and that functional connectivity between the subgenual ACC and MTL was enhanced in MDD. Kwaasteniet also identified the negative correlation between uncinate fasciculus integrity and subgenual ACC functional connectivity with the bilateral hippocampus in MDD but not in healthy controls; this negative structure-function relation was positively associated with depression severity (Kwaasteniet et al., 2013). Zhang and his colleagues (2011) published a fMRI/graph theory (small world) study confirming disrupted brain connectivity networks in drug-naïve first-episode MDD. It seems that MDD disrupts the global topological organization of the whole-brain networks. There are studies that emphasized the disrupted brain connectivity in mental disorders (van Essen et al., 2012; Castellanos et al., 2013; Kim et al., 2013).

Lee et al. (2011) tested the connectivity strength of resting state EEG as a potential biomarker of treatment response in major depressive disorder. They concluded that '…the stronger the connectivity strengths, the poorer the treatment response.' The experiment also showed that frontotemporal connectivity strengths could be a potential biomarker to differentiate responders from slow responders and non-responders in MDD. Chen et al (2017) reported higher amplitude of low-frequency fluctuations (ALFF) in both the amygdala and hippocampus in participants with MDD compared to their healthy peers. Using graph theoretical analysis, they found that clustering coefficient, local efficiency, and transitivity are decreased in MDD patients (Chen et al., 2017). In their reviews, Drevets et al. (2008) and Willner et al. (2013) covered almost all aspects examined in the quest of understanding the characteristic features of depression. The first one focused on structural and functional abnormalities and neurocircuitry in depression, and the second one more broadly reviewed the present and dominant approaches in this area of research. Prior to this review, Willner et al. (2005) also demonstrated that antidepressants do not normalize brain activity: 'mood and behavior are restored to normal, but antidepressant-treated brain is in a different state from the non-depressed brain' (Willner et al., 2005). As a sum, all the changes found in depression indicate that the main characteristic of MDD is actually in their abnormal connectivity and transfer of information, rather than in solely physical differences. In their Granger Causality study about depression, Hamilton and his colleagues (Hamilton et al., 2011) are questioning the importance of the functional connections between candidate regions found to be abnormal in depression. Their research relied on then handful number of prior studies that yielded information about cross-structural communication and influence in depression (Lozano et al., 2004; Seminowicz et al., 2008). Based on previously confirmed aberrant interrelations in MDD, they applied multivariate Granger Causality to estimate the extent to which preceding neural activity in one or more seed regions predicted subsequent activity in target brain regions in the analysis of blood oxygen-level-dependent (BOLD) data.

Hamilton et al (2011) found that increased activity in ventral anterior cingulate cortex (vACC) could be predicted by the activation of hippocampus in patients with depression. In addition, the authors showed a mutual reinforcing effect between vACC and prefrontal cortex. Further, it was found that hypoactivity of dorsal cortical regions might be predicted with vACC and hippocampal increased activity.

They demonstrated that aberrant patterns of effective connectivity implicate disturbances in the mesostriatal dopamine system in depression contributing to the knowledge about the primary role of limbic inhibition of dorsal cortex in the cortico-limbic relation (Hamilton et al., 2011). It seems that many above mentioned areas probably illustrate different dynamics as networks, active in specific tasks known to be characteristically different in depression.

The aim of this paper is to compare the network dynamics of the MDD and healthy brain applying Transfer Entropy (TE) to quantify the directed dynamical interactions in the resting state electroencephalographic (EEG) data. For doing this we used MuTe MATLAB Toolbox as a freeware designed to evaluate transfer entropy (TE) which is able to quantify the directed dynamical interactions (Montalto et al., 2014). To map information transfer between some of already mentioned structures confirmed in former EEG research, we developed an algorithm in Java programming language, based on previously published MuTe MATLAB Toolbox (Montalto et al., 2014), and applied it on resting-state EEG recordings from participants diagnosed with MDD and a control group.

**Methods**

*Mathematical Tools*

Transfer Entropy (TE) is well based on information theory and is model-free, which makes it sensitive to all types of interactions between time series under study. MuTe MATLAB Toolbox is a freeware designed to evaluate transfer entropy (TE) which is able to quantify the directed dynamical interactions (Montalto et al., 2014). To map information transfer between some of already mentioned structures confirmed in former EEG research, we developed an algorithm in Java programming language, based on previously published MuTe MATLAB Toolbox (Montalto et al., 2014), on resting-state EEG recordings from participants diagnosed with MDD and a control group.

In this section we will briefly describe the methods that we used to assess the directed dynamical links among the recorded time series.

When we observe a complex system which consists of M interacting dynamical subsystems and we want to evaluate an information flow from the source system X to the destination system Y, first we describe the vector $Z_k=1,\ldots,M-2$, for all the remaining systems. This framework was originally developed under the assumption of stationarity. That allows us to do the estimation by replacing ensemble averages with time averages. We are denoting *X, Y* and *Z* as stationary stochastic processes. *X,Y* and *Z* are described by the states which were visited by the systems over the time, and the stochastic variables $X_n$, $Y_n$ and $Z_n$ which were obtained by the sampling the processes at the present time *n*. Further, we denote:

$$X_n^- = [X_{n-1} X_{n-2} \ldots],\ Y_n^- = [Y_{n-1} Y_{n-2} \ldots],\ \text{and}\ Z_n^- = [Z_{n-1} Z_{n-2} \ldots] \tag{1}$$

where the vector variables are representing the past of mentioned processes *X*, *Y* and *Z*. To take into account the instantaneous influences of the candidate drivers in some cases is also recommendable. In this case, the vectors *Xn*– and *Z*–*n* defined above should contain the present terms *Xn* and *Zn* too. The multivariate transfer entropy from *X* to *Y* (which is conditioned by *Z*) can be defined as:

$$TE_{X \to Y|Z} = \sum p\left(Y_n, Y_n^-, X_n^-, Z_n^-\right) \log \frac{p\left(Y_n | Y_n^-, X_n^-, Z_n^-\right)}{p\left(Y_n | Y_n^-, Z_n^-\right)} \quad (2)$$

As the sum extends over all the phase spaces points thus forming the trajectory of the complex (composite) system. *p*(a) is then the probability associated with the vector variable a, while *p*(*b*|a) = *p*(a,*b*)/*p*(a) is the probability of observing *b* knowing the values of a.

Since TE does not assume any particular model which is describing the interactions behind the systems dynamics, it has a great potential in information transfer detection; TE is able to discover purely non-linear interactions and to deal with a range of interaction delays (Vicente et al., 2011). For the data that can be assumed to be drawn from a Gaussian distribution it is shown that TE is equivalent to Granger Causality (GC); the data covariance is fully described by a linear parametric model (Barnett, 2009; Hlavačkova-Schindler et al., 2011). This establishes a convenient joint framework for both measures. In this work we are evaluating GC in the TE framework. Also we aim at comparing this approach to a model-free approaches. The TE estimator is the binning estimator (BIN) previously described in (Montalto et al., 2014). It consists of coarse-graining of the observed dynamics, by utilizing Q quantization levels, after which the entropies are computed by approximating probability distributions with the frequencies of the occurrence of the values quantized (Hlavačkova-Schindler, 2011). Referring to the MuTE toolbox (Montalto et al., 2014), we used the BINNUE, the binning estimator in the framework of the non-uniform embedding.

From a set of candidate variables (including past of *X*, *Y* and *Z*) a progressive selection leads to a non-uniform embedding approach. Here the past of all candidate variables was considered up to a maximum lag (candidate set) of the lagged variables since they are the most informative for the target variable $Y_n$. The selection is performed at each step by maximizing the amount of information that can be explained by observing variables considered with their specific lag up to the current step. This results in a criterion for the maximum relevance and minimum redundancy for candidate selection, so that the resulting embedding vector $V = [V_{nX} V_{nY} V_{nZ}]$ includes only the components of $X_{n-}$, $Y_{n-}$ and $Z_{-n}$, which contribute most to the description of $Y_n$.

We implement a growing neural network to study dynamical interactions in a system made up of several variables, described by time series. The aim is not only to find a directional relationship of influence between a subset of time series, the *source*, and a *target* time series taking into account the rest of the series collected in a set, called *conditioning*, but also to determine the delay at which the source variables are influencing the target.

For directed dynamical influences among variables, modeled as time series, neural networks were used as a powerful tool to compute the prediction errors needed to evaluate causality in the Granger sense. According to the original definition, Granger causality (GC) deals with two linear models of the present state of a target variable. The first model does not include information about the past states of a driver variable, while the second model does. If the second model's error is less than that of the first model in predicting the present state of the target, then we can safely say that the driver is causing the target in the sense of Granger (Granger, 1969). Here we introduce a new Granger causality measure called *Neural Networks Granger Causality (NNGC)* defined as

$$NNGC = err_{reduced} - err_{full} \qquad (3)$$

where $err_{reduced}$ is the prediction error obtained by the network that does not take into account the driver's past states, while $err_{full}$ is the prediction error evaluated by the network that takes into account the driver's past states.

To better explain the target series, we used the non-uniform embedding technique. Instead of fitting linear models originally proposed by Granger (predefined models), here we train a neural network to estimate the target using the past states only. Such strategy leads to growing neural networks, with an increasing number of input neurons, each input neuron representing a past state chosen from the amount of past states available, considering all the variables in the system. The present approach combines non-uniform embedding and a regularization strategy by a validation set to detect dynamic causal links. The validation phase is then embedded in the learning phase: this combination of training and validation avoids erroneous use of the training procedure, thus avoiding overfitting.

*Participants*

In this study we re-used EEG dataset recorded for previously performed study (Čukić et al., 2018; 2019; 2020). Participants with MDD were recorded at the Institute for Mental Health, Belgrade, Serbia, and healthy control group (HC) was recorded at the Institute for Experimental Phonetic and Speech Pathology, Belgrade, Serbia. MDD group comprised 21 participants (nine male), 25 to 68 years (32.4±10.16). The HC group comprised 20 participants (10 male) matched in age (30.14±8.94). The participants from HC group had no history of psychiatric or neurological disorders. The participants were right-handed according to Edinburgh Handedness Inventory. Each participant gave his/her written informed consent prior the EEG recording. The entire experimental protocol was approved by the Ethical Committee of the Institute for Mental Health (Approval number 30/59-27/10/2015) and the Institute for Experimental Phonetic and Speech Pathology (Approval number 87-EO/15-25/9/2015) according to Declaration of Helsinki regarding experiments involving human participants. All participants with depression were on medications and under the supervision of an experienced clinical psychiatrist. Their diagnoses were made according to ICD-10.

*Data acquisition*

The participant's EEG was recorded during eyes closed condition, in a sitting position. All recordings were performed around noon (±1 hour). The electrodes were positioned according to 10/20 International system for electrode placement. The MDD group was recorded using Nicolet One Digital EEG Amplifier (VIAYSYS Healthcare Inc. NeuroCare Group) apparatus with 19 electrodes implemented in cap (Electro-cap International Inc. Eaton, OH USA). The montage was monopolar. The sampling rate was 1000Hz. The resistance was kept less than five KOhm. A bandpass filter was 0.5-70 Hz. The same experimental setup was used for the EEG recording of the HC group but with Nihon Kohden Inc apparatus. Although the EEG recording was performed on two different apparatus but with the same setup, there were no differences between groups (Pivik et al., 1993). Further, we used records from 21 patients and 20 healthy controls for this study. Artifacts were carefully inspected visually and marked manually in order to be avoided from further analyses. Each recording session lasted for 5 minutes. Each epoch for further analysis comprised 2000 samples.

**Results**

The aim of this analysis was to probe BINNUE approach in order to characterize the dynamical interactions between standardized areas in patients diagnosed with depression in contrast to healthy controls. We were looking at the information transfer between the standardized positions in 10/20 EEG cap (corresponding to the records from 19 channels) and the detected directions. The results are reported in Figure 1, showing the averaged results for 21 MDD and for 20 healthy controls.

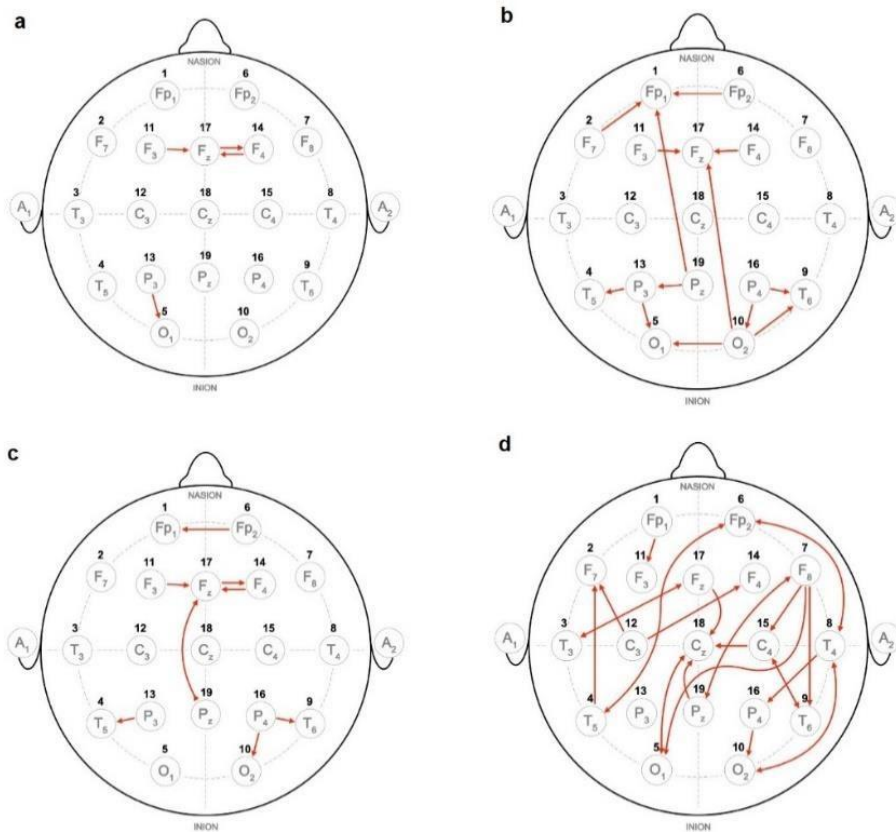

**Figure 1**: Results for BINNUE analysis. The figures a, b and c represent the analysis performed on segments 1, 2 and 3, respectively, taken from 5-minute EEG recording of Patient group (those with MDD) with the standard 10/20 system. Figure d is a representation of BINNUE results for Healthy control group (HC). Standard names of electrode positions according to 10/20 system are Fp1 to Pz, but the bold number above (1-19) corresponds to the order of the channels analyzed in our analysis.

The BINNUE approach provided us with both influence and directions of influence between compared time series (epoch extracted from recorded EEG). In figure 1 (d) we can observe seven two-directional arrows implying that the influence was in both directions. Those are: Fp2-T5, Fp2-T4, T3-Fz, Cz-O1, Pz-F8, C4-T6 and T4-O2. Other detected influences are in one direction, namely: Fp1-F3, C3-F7, T5-F7, C3-F4, Fz-Cz, Pz-Cz, P4-O2, T4-P4, F8-C4, F8-T6.

In all three figures representing influential connections in MDD group, we have much less registered connections. From segment 1 (Fig1/a) two one-directional connections (F3-Fz and P3-O1) and just one bi-directional (Fz-F4). From segment 2 (Fig1/b) there are only one-directional influences (13 of them): Very strong Fp2-Fp1, O2-Pz and Pz-Fp1, and les strong Fz-Fp1, F3-Fz, F4-Fz, O2-P1, O2-T6, P4-O2, P4-T6, P3-T5, P3-O1 and Pz-P3.

In Fig1/c we can see the results calculated from segment 3 (MDD). There are six one-directional (Fp2-Fp1 very strong) and F3-Fz, Pz-Fz, P3-T5, P4-O2 and P4-T6. Only one of detected connections is exhibiting very strong bi-directional influence: Fz-F4.

From BINNUE analysis it is obvious that the dynamics of information transfer differs between healthy controls group and MDD participants.

**Discussion**

According to our results, the dynamics of healthy resting-state EEG is substantially different from the dynamics of MDD brain: the interactions (information transfers) in healthy controls are numerous during resting state, contrary to MDD brains which are repeatedly showing the "isolated" activity in frontal, parietal and temporal areas. It should be emphasized that F3 (left DLPF)-Fz (frontal midline region)-F4 (right DLPFC) regions show engagement in each segment. Knowing that right DLPFC region is involved in processing negative emotions (fear, anxiety, sadness), our results are in line with the previously described inability of persons with MDD to disengage from negative emotional content, as described in Willner (2013), and Berman et al. (2011), and Gotlieb and Joorman (2010). Graphical representation of transfer entropy (TE) in MDD shows just a few nodes (electrode positions) probably representing the abnormal cortical functional connectivity as a reflection of the one within the fronto-limbic system (Vederine et al., 2010; de Kwaasteniet et al., 2013), engaged in negative information.

Recent research showed that TE is equivalent to Granger Causality (GC) for the data drawn from a Gaussian distribution (Barnett et al., 2009; Hlavačkova-Schindler et al., 2011). TE can be seen as a difference of two conditional entropies (Montalto et al., 2014) and can detect information transfer, discovering purely non-linear interactions between time-series under study. When we compare our findings with previously used GC results (on BOLD dataset see Hamilton et al., 2010), there are a certain number of connections that are in line with those findings. Hamilton et al. (2010) found that while observing moment to moment interactions, hippocampi were influencing vACC and consequently decreased activation of DLPFC. It seems that hippocampus has a critical role in affecting depresotypic neural responses (Ge et al., 2019). Of course, we cannot claim that we detected anything below the level of cortex by EEG, but the connections from Cz-Fp1 (Fig2/a), Pz-Fp1 (Fig1/b), and Pz-F2 (Fig1/c) are illustrating the direction of influences. It seems that F3, Fz and F4 are the most pronounced way of the information flow in our results.

It is also important to note that we chose to analyze resting-state EEG due to former results of other researchers. First of all, Goldberger repeatedly showed (Goldberger et al., 2002; 2006) that resting state is the most information-rich, and should be considered of the utmost importance when we want to learn about healthy functioning of complex systems. Berman et al. (2011) also showed that rumination can be related to neural signaling only in task-free states. We think that employing an elaborate measure using entropy yields valid results that can be properly interpreted.

From the present literature in the field of nonlinear analysis of EEG, we can conclude that a certain consensus exists about the elevated complexity in the signal in persons diagnosed with the depression (de la Torre-Luque & Bornas, 2017). Researchers applied very different algorithms in a mathematical sense (fractal, entropy measures, Detrended Fluctuation Analysis, Quantitative Recurrent Analysis, Geometric measures, etc.) but all of them consistently reported higher levels of complexity in depressed patients in comparisons to controls (Ahmadlou et al., 2011; Faust et al., 2014; Hosseinifard et al., 2014; Bachmann et al., 2015; 2018, Lebliecka et al., 2018; Jaworska et al., 2018; Čukić et al., 2019; 2020). Since we succeeded in illustrating purely non-linear interactions between time series recorded from standard positions which have anatomical meaning, it is possible to relate our work to the results of other researchers that used a quite different methodology to examine their data. Hence, we can say that the methodology we used is capable of detecting the variables representing the past of the process, implying the causality between recorded signals in a very straightforward way.

BINNUE successfully revealed connections involving DLPFC, based on nonlinear measures calculated from raw EEG as features for various machine learning methods (Čukić et al., 2019;

2020). Our results suggest that the use of model-free TE estimators in our work for detecting the information transfer in physiological time series is highly justified.

Once we have the tool to examine which series influenced each other in the process, like the one we already know to be aberated in depression, further research on much larger sample is a logical step (since our results cannot be properly generalized due to a modest sample). We strongly believe that TE is this tool and that it can help us broaden our understanding of underlying physiological processes. To the best of our knowledge, this is the first graphical representation of information transfer in MDD and healthy controls and it can be seen how it differs between the two. The BINNUE approach provided us with both influence and directions of influence between compared time series (epochs extracted from recorded EEG)

**Funding**: S.R. is not granted for this work, M.Č. was granted by Project H2020-MSCA-RISE-2015-690874 (2016-2020). D.S. was granted by Serbian Ministry of Education, Science and Technology 2018, number III41029, M.S. was granted by Serbian Ministry of Education, Science and Technology OI178027 and TR32032.

**Contributions**: M.Č. conceived and designed experiment, adapted, cleaned and choose the epochs for further analysis. M.Č. and S.R. performed analyses and made graphical representations, S.R. wrote algorithm in Java. M.Č. and D.S. interpreted the results and wrote the manuscript. M.S. recorded and sequenced EEG signal, and took part in reviewing the manuscript.

**Conflict of interest**: Authors have no conflict of interest to report.